\documentclass[twocolumn,           
               showpacs,            
               preprintnumbers,     
               aps,                 
               prl,                 
               a4paper,             
               superscriptaddress,  
               nofootinbib,         
               tightenlines,        
               floats,floatfix      
               ]{revtex4}

\usepackage{graphicx}
\usepackage{dcolumn}
\usepackage{bm}
\usepackage{latexsym}
\usepackage{amsmath,amssymb}        

\begin{document}
\title{Leptogenesis and Reheating in Complex Hybrid Inflation}
\author{Carlos Mart\'inez-Prieto}
\email{carlosr@ifm.umich.mx}
\affiliation{Instituto de F\'isica y Matem\'aticas de la Universidad
  Michoacana de San Nicol\'as de Hidalgo, Edificio C-3,
  Cd. Universitaria, A.P. 2-82, 58040, Morelia, Michoac\'an, M\'exico.}
\author{David Delepine}
\email{delepine@fisica.ugto.mx}
\author{L. Arturo Ure\~na-L\'opez}
\email{lurena@fisica.ugto.mx}
\affiliation{Departamento de F\'isica, DCI, Campus Le\'on, Universidad
  de Guanajuato, C.P. 37150, Le\'on, Guanajuato, M\'exico.}

\date{\today}

\begin{abstract}
We study the transformation into a baryon asymmetry of a charge
initially stored in a complex (waterfall) scalar field at the end of a
hybrid inflation phase as described in Ref.~\cite{Delepine:2006rn}. The
waterfall field is coupled to right-handed neutrinos, and is also
responsible for their Majorana masses. The charge is finally
transferred to the leptons of the Standard Model through the decay of
the right-handed neutrinos without introducing new CP violating
interactions. Other needed processes, like the decay of the inflaton
field and the reheating of the Universe are also discussed in detail.
\end{abstract}

\pacs{98.80.-k; 98.80.Cq; 98.80.Ft}

\maketitle

\section{Introduction \label{sec:introduction-}}
We have shown in a recent paper\cite{Delepine:2006rn} that a complex
hybrid inflation model can generate a charge asymmetry that may be
further transferred into a baryon charge, then providing a possible
solution to the baryogenesis problem of Cosmology\cite{Kolb:1990vq}.

The scalar field potential associated of the complex hybrid inflation
model is
\begin{eqnarray}
  V(\phi,a) &=& \frac{1}{4\lambda^2} (M^2 - \lambda^2 \vert a \vert^2
  )^2 + \left( \frac{m^2}{2} + \frac{g^2}{2} \vert a \vert^2 \right)
  \phi^2 \nonumber \\
  && + \frac{\delta}{4} a^2 \phi^2 + \mathrm{c.c.} \label{eq:1}
\end{eqnarray}
where $\phi$ is the inflaton field and $a$ is the (complex) waterfall
field. Notice that there is an explicit term that violates the $U(1)$
global symmetry. The needed charge is generated during inflation and
is associated to the charge of the waterfall field $a$ that puts an end to
inflation.

The aim of this paper is to discuss the transfer of the $a$-charge to
fermionic matter after the end of inflation. The task is not a simple
one, there may be some transfer processes during the reheating phase
of the Universe that can wash out the generated charge.

To fix ideas, we shall work on a leptogenesis model in which the
$a$-charge is transferred to Standard Model particles through
interactions between the waterfall field $a$ and a right-handed
neutrino $N_R$. The interaction Lagrangian reads
\begin{equation}
  \mathcal{L} = h_{Y} \bar{\ell}_{L} \Phi N_{R} + h_{2} \bar{N}_{R}^c
  N_{R}a + \mathrm{h.c.} \label{eqs:2}
\end{equation}
where $\Phi$ is the Higgs doublet, $\ell_{L}$ is the leptonic doublet,
and $h_{Y}$ is the usual Yukawa coupling.

The interaction Lagrangian~(\ref{eqs:2}) is inspired in the majoron
model and in the standard leptogenesis
scenario\cite{Fukugita:1986hr,Luty:1992un,Riotto:1998bt,Frere:2008ct}. Our
model is then composed of just one family of leptons that contains one
leptonic doublet and one right handed neutrino.

A summary of the paper is as follows. In Sec. we briefly review the
inflationary dynamics of the complex hybrid inflation model as
presented in Ref.\cite{Delepine:2006rn}. In Sec. we study the
post-inflationary dynamics of the different fields involved in the
model, and focus our attention in the stages of preheating and
reheating that may appear. Sec. is entirely devoted to the study of
the Boltzmann equations in order to estimate the amount of the
$a$-charge that is finally converted into a useful baryon
charge. Finally, conclusions are presented in Sec.

\section{Inflationary dynamics \label{sec:infl-dynam-}}
The model is given by the potential~(\ref{eq:1}), where $g$ and
$\lambda$ are real constants, and $\delta$ is a complex
parameter; for $\delta=0$ we recover the standard hybrid inflation
model\cite{Felder:2000hj}. The $\delta$-term violates the $U(1)$
symmetry associated to the complex field $a$. However, the
potential~(\ref{eq:1}) is $CP$ conserving as the phase of complex
parameter $\delta$ can be removed through a phase redefinition of the
$a$-field.

The scalar potential has a local maximum at $\phi=\vert a\vert =0$
with a height given by $V(0,0)= M^4/(4\lambda^2)$ that corresponds to
a false vacuum. The true vacuum of the system corresponds to the
global minimum located at $\phi=0$ and $\lambda \vert a \vert
/M=1$; this true vacuum is degenerate. The $U(1)$ charge density at
any time is given by $n_a=a_r\dot{a}_i-a_i\dot{a}_r$ where the r and i
refers to the real and imaginary components of the $a$ field.

The constant term in the potential ~(\ref{eq:1}) is initially the
dominant one, which is usually dubbed as false vacuum
inflation\cite{Copeland:1994vg}. In the regime of slow-roll the scale
factor grows exponentially with time, $R(t) = R_{end} \exp[ H_0 (t -
t_{end})]$, whereas the evolution of the inflaton field is given by
$\phi(t) =\phi_{end} \exp[( m^2/3H_0 )( t_{end} -t )]$, where $H_0$ is the
(almost constant) Hubble parameter during inflation.

The waterfall field is trapped in the false vacuum  $a=0$, but when
the inflaton field passes through the value $\phi_- = M/\sqrt{g^2 -
  \delta}$ the imaginary component of the waterfall field $a_i$
presents a tachyonic instability\cite{Felder:2000hj}, and falls down
towards its true vacuum value. Likewise, when the inflaton field
passes through the value $\phi_+ = M/\sqrt{g^2 + \delta}$ the real
component $a_r$ becomes unstable and moves too to its true vacuum
value; it is at this point that inflation ends.

The asymmetric evolution of the components of the waterfall field
generates a dynamical $CP$ violating phase during the phase transition
at the end of inflation, and produces an asymmetry in the charge of
the $a$ field.

Taking into account different observational constraints, it was shown
that the appropriate values of the parameters correspond to the
case $\lambda^2 \gg g^2 \gg \delta$. Hereafter, we will consider these
to be the right case for the parameters of our model.

\section{Post-inflationary dynamics and
  reheating \label{sec:post-infl-dynam}}
We begin at the end of the inflation once the inflaton field passes
through the second critical point $\phi_+$. The dynamics afterward
depends upon the values of the different parameters in the model, but
it very much resembles that of typical hybrid inflation coupled to a
third massless scalar field. We shall follow the calculations
presented in
Refs.\cite{Kofman:1994rk,Kofman:1997yn,GarciaBellido:1997wm}, where
more details can be found.

\subsection{Dynamics of preheating \label{sec:dynamics-preheating-}}
First, we revisit the critical points of the potential (\ref{eq:1});
they are to be found from the equations
\begin{subequations}
  \begin{eqnarray}
    \frac{\partial V}{\partial \phi} &=& \left[ m^2 + g^2 |a|^2 +
      \delta (a^2_r - a^2_i ) \right] \phi = 0 \, , \label{eq:1a} \\
    \frac{\partial V}{\partial a_r} &=& \left[ -M^2 + \lambda^2 |a|^2
      + \phi^2 (g^2+\delta) \right] a_r = 0 \,, \label{eq:1b} \\
    \frac{\partial V}{\partial a_i} &=& \left[ -M^2 + \lambda^2 |a|^2
      + \phi^2 (g^2-\delta) \right] a_i = 0 \, . \label{eq:1c}
  \end{eqnarray}
\end{subequations}
As we mentioned before, the critical points exist only for $\phi=0$;
they are the origin of coordinates, $(\phi=0, |a|=0)$, and the
degenerate circle on the complex plane, $(\phi=0, |a|=M/\lambda)$.

\begin{figure*}[!htp]
  \includegraphics[width=8.7cm,height=8cm]{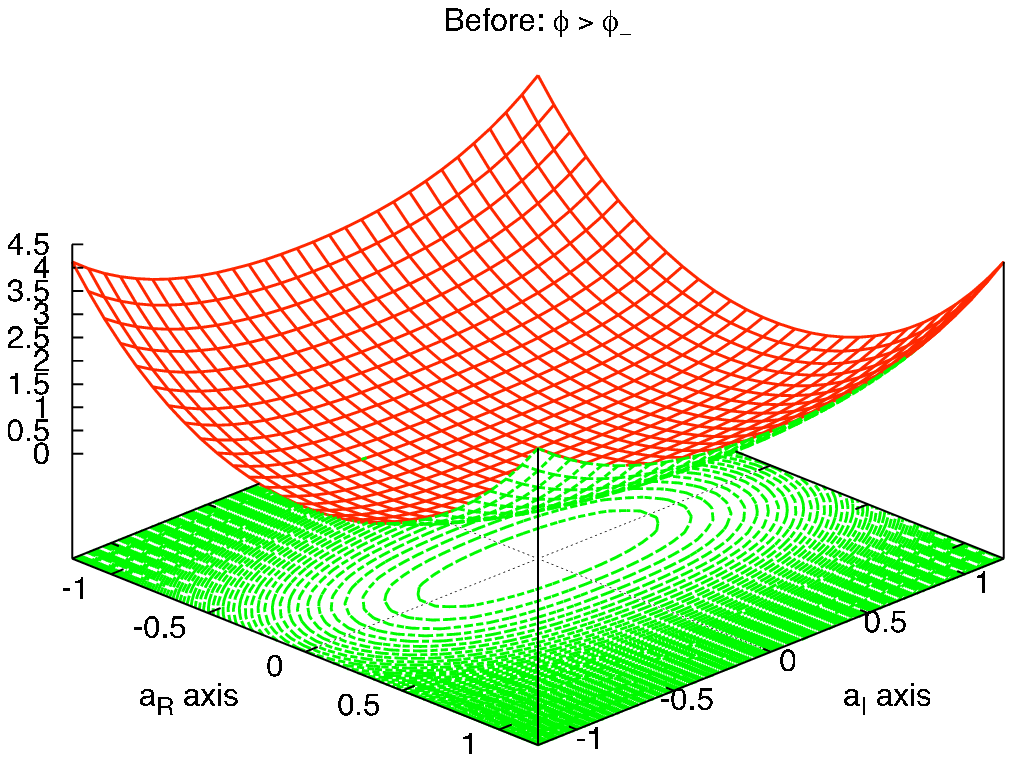}
  \includegraphics[width=8.7cm,height=8cm]{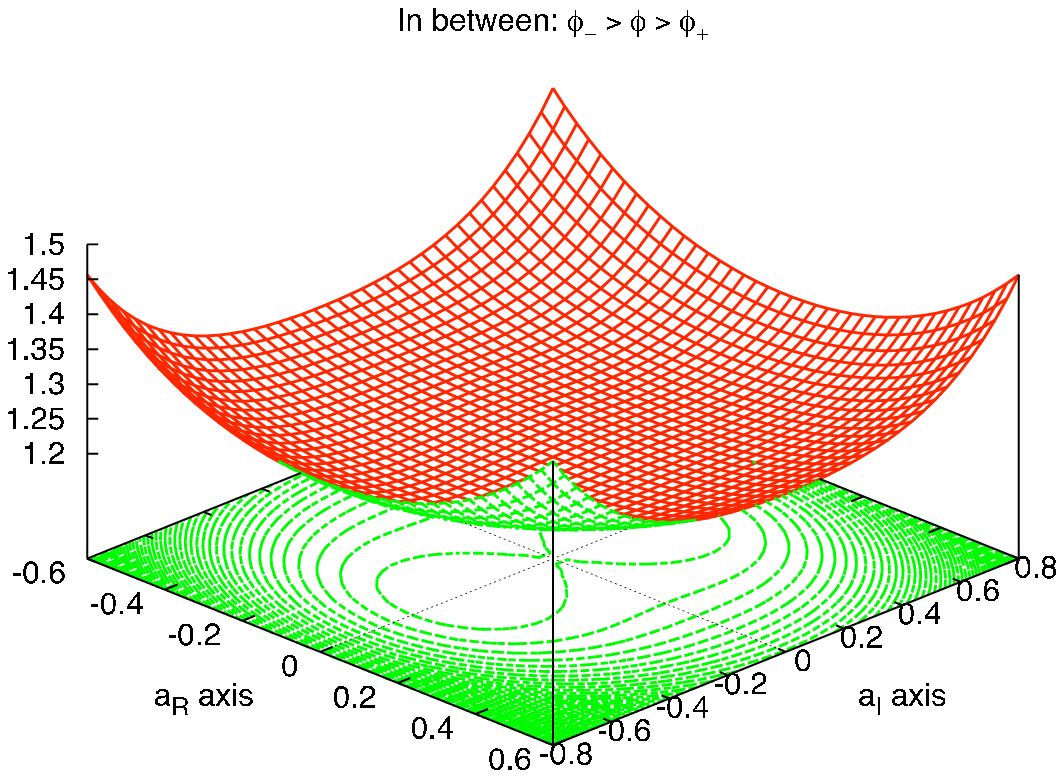}
  \includegraphics[width=8.7cm,height=8cm]{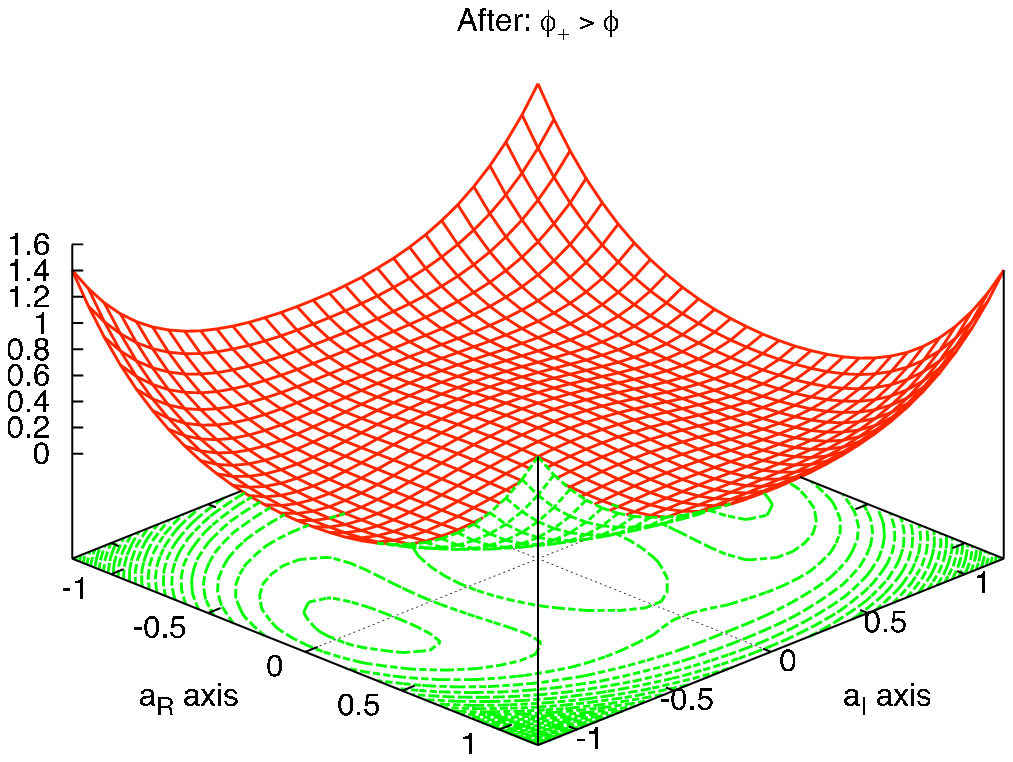}
  \includegraphics[width=8.7cm,height=8cm]{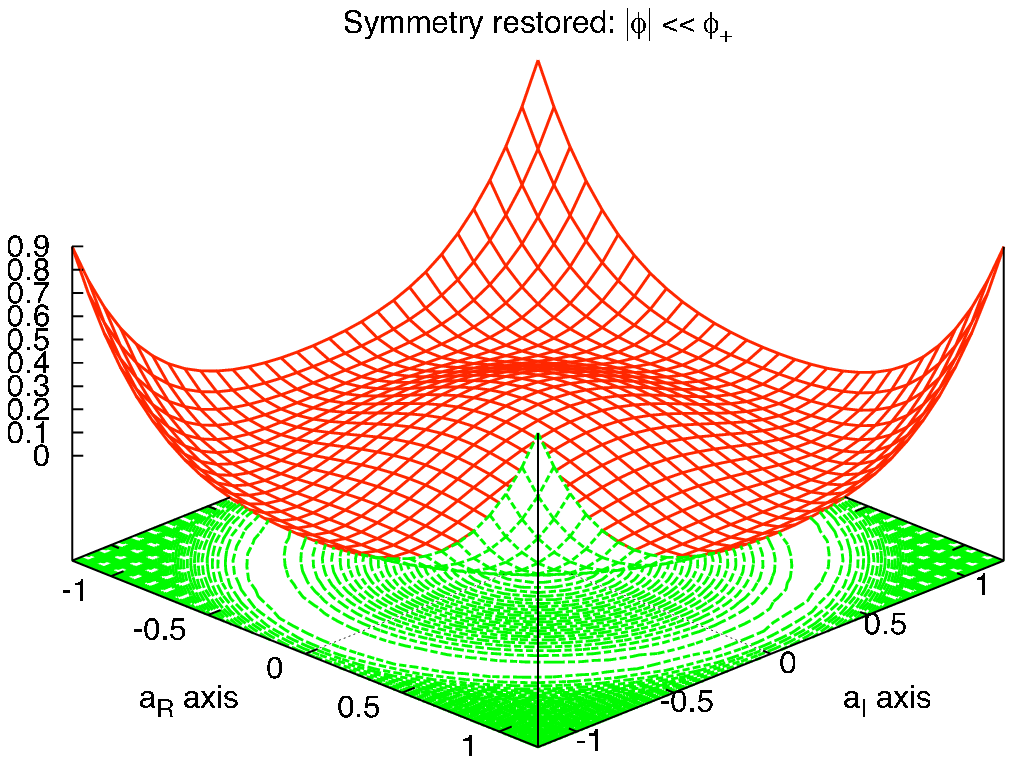}
  \caption{\label{fig:1} Different snapshots of the
    potential~(\ref{eq:1}) as projected on the complex plane of the
    waterfall field $a$ and for different values of the inflaton field
    $\phi$. There is only one minimum during inflation, but two minima
  appear after the crossing of the first instability point $\phi_-$;
  the full restoration of the $U(1)$ symmetry is almost complete well
  after the end of inflation.}
\end{figure*}

However, it is very instructive to consider the position of the
critical points on the complex plane $(a_r,a_i)$ for non-zero values
of the inflation field. We notice that the location of the critical
points changes with time as shown in Fig.~\ref{fig:1}. There are 5
critical points after passing through the first instability point
$\phi_-$, but only two of them correspond to minima of the scalar
potential; in general, the two minima are located along the real axis,
$a_r = 0$. The critical values of the imaginary part of the waterfall
field are determined from the ellipse equation
\begin{equation}
  \label{eq:2}
  \lambda^2 a^2_i + (g^2-\delta) \phi^2 = M^2 \, .
\end{equation}

On the other hand, we should take into account the effective masses of
the different fields; they are given by
\begin{subequations}
  \begin{eqnarray}
    m^2_{\phi} &=& \frac{\partial^2V}{\partial \phi^2} = m^2 + g^2
    |a|^2 + \delta (a^2_r - a^2_i ) \, , \label{eq:2a} \\
    m^2_{a_r} &=& \frac{\partial^2V}{\partial a^2_r} = -M^2 + \lambda^2
    |a|^2 + 2 \lambda^2 a^2_r + (g^2 + \delta) \phi^2 \,
    , \label{eq:2b} \\
    m^2_{a_i} &=& \frac{\partial^2V}{\partial a^2_i} = -M^2 + \lambda^2
    |a|^2 + 2\lambda^2 a^2_i + (g^2 - \delta) \phi^2 \,
    . \label{eq:2c}
  \end{eqnarray}
\end{subequations}
The values of the masses at the effective minima can be rewritten as
\begin{eqnarray}
  m^2_\phi &=& m^2 + \frac{(g^2 - \delta)}{\lambda^2} M^2 \left( 1 -
    \frac{\phi^2}{\phi^2_-} \right) \, , \\
  m^2_{a_r} &=& 2 \delta \phi^2 = \frac{2\delta}{g^2 + \delta}
  \left( \frac{\phi^2}{\phi^2_-} \right) \, , \\
  m^2_{a_i} &=& 2 M^2 \left(1 - \frac{\phi^2}{\phi^2_-} \right) \, .
\end{eqnarray}
where $\phi_-$ is the first instability point at the end of
inflation. We notice that the field $a_r$ is practically massless, and
in the limit $\phi \to 0$ we find $m_\phi \to \bar{m}_\phi \equiv g M
/ \lambda$, $m_{a_r} \to 0$, and $m_{a_i} \to \bar{m}_{a_i} \equiv
\sqrt{2} M$.

From the discussion above, we see that as the system approaches the
true vacuum of the system the $3$-fields system we are dealing with
can be matched to the standard theory of hybrid inflation plus the
addition of a massless field. That is, our system $(\phi,a_i,a_r)$
behaves as in the theory of preheating in hybrid inflation presented
in Ref.\cite{GarciaBellido:1997wm}, in which the three fields in
interaction are $(\phi,\sigma,\chi)$ being $\chi$ the massless
field. Because of this similarity, our next calculations follow those
presented in Ref.\cite{GarciaBellido:1997wm}.

For the dynamics after inflation, two different regimes have been
identified. For reasons explained before, we shall be interested in
the regime corresponding to $\lambda^2 \gg g^2$.  The field $a_r$
behaves as a massless field, and we expect it to evolve adiabatically
following the instantaneous position of its critical point at $a_r =
0$.

As for the other fields, $\phi$ and~$a_i$, they also evolve
adiabatically along the ellipse~(\ref{eq:2}); the field $a_i$ does it
with negligible oscillations and then most of the energy is stored in
the oscillations of the inflaton field.

Next step is to study the production of $\phi$ and $a$-particles
during the so-called preheating stage which happens during the
oscillating phase of the fields; for this we need to write the
evolution equations of the quantum fluctuations $\delta \phi_k$,
$\delta a_{rk}$, and $\delta a_{rk}$ in linear perturbation theory.

It is better to define new variables $\varphi = R^{3/2} \delta \phi$,
$\psi = R^{3/2} \delta a_i$, and $\eta = R^{3/2} \delta a_r$; thus,
the production rate of particles of the different fields is determined
by the Mathieu equations
\begin{subequations}
\begin{eqnarray}
  \varphi^{\prime \prime}_k + \left[ A_\phi(k) - 2q_\phi \cos(2z)
  \right] \varphi_k &=& 0 \, , \\
  \psi^{\prime \prime}_k + \left[ A_{a_i}(k) - 2q_\psi \cos(2z)
  \right] \psi_k &=& 0 \, , \\
  \eta^{\prime \prime}_k + \left[ A_{a_r}(k) - 2q_\eta \cos(2z)
  \right] \eta_k &=& 0 \, ,
\end{eqnarray}
\end{subequations}
where a prime denotes derivative with respect to $z=m_\phi
t$. The production of particles can be specially efficient if $A(k)
\ll 2q$ and $q \geq 1/4$, and then it is necessary to study each case
separately.

\textbf{$\phi$-particles}. The Mathieu parameters are
\begin{equation}
  A_\phi(k) = \frac{k^2}{R^2 \bar{m}_\phi} + 1 + 2q_\phi \, , \quad
  q_\phi = \frac{\Phi^2(t)}{4} \, ,
\end{equation}
where $\Phi^2(t) = 2 \langle \phi^2/\phi^2_- \rangle \propto 1/t^2$ is
the averaged squared amplitude of the inflaton oscillations around the
minimum of the potential, and $R(t)$ is the scale factor of the
Universe.

For the regime $\lambda^2 \gg g^2$, it is possible that the inflation
field performs large-amplitude oscillations after inflation, and then
initially $q_\phi \simeq 1/4$; this may indicate a successful scenario
for the production of particles. However, careful studies show that
the small oscillations of the field $a_i$ prevent the existence of an
explosive production of $\phi$-particles\cite{GarciaBellido:1997wm}.

\textbf{$a_i$-particles}.  The Mathieu parameters are
\begin{equation}
  A_{a_i}(k) = \frac{k^2}{R^2 \bar{m}_\phi} + \frac{2\lambda^2}{g^2} +
  2q_{a_i} \, , \quad q_{a_i} = \frac{2 \lambda^2}{g^2}
  \frac{\Phi^2(t)}{4} \, .
\end{equation}
Even though $q_{a_i} > 1/4$ initially because of the regime $\lambda^2
\gg g^2$, this also means that $A_{a_i} \gg q_{a_i}$. Then, our system
is far above the resonance band for the production of particles; thus,
no explosive production of $a_i$-particles is expected.

\textbf{$a_r$-particles}.  The Mathieu parameters are
\begin{equation}
  A_{a_r}(k) = \frac{k^2}{R^2 \bar{m}_\phi} + 2q_{a_i} \, , \quad
  q_{a_r} = \frac{2 \delta}{g^2} \frac{M^2}{\bar{m}^2_\phi}
  \frac{\Phi^2(t)}{4} \, .
\end{equation}
Initially, we would find that $q_{a_r} \simeq \lambda^2 \delta / g^4
\ll 1$; thus, we again find that we cannot expect an explosive
production of $a_r$-particles.

In consequence, all the considerations above strongly suggest that the
dynamics of the complex hybrid inflation model is (almost) entirely
described by the classical evolution of the inflaton and waterfall
fields. In other words, there are not important preheating processes
for the production of particles after inflation, and we can say that
all fields evolve coherently in very good approximation.

\subsection{Reheating}
Now that we have established the coherence of the inflaton
oscillations, we must consider the reheating of the Universe after
inflation; that is, the transfer of the inflaton's energy into
relativistic degrees of freedom in its initial phase of rapid
oscillations \cite{Podolsky:2005bw}.

The (perturbative) theory of reheating says that the inflaton can
decay into other (whether scalar or fermionic) degrees of freedom if
its mass is larger than those of the products. In the case of interest
here, $\lambda^2 \gg g^2$, the effective mass of the $a_i$-field is
$m_{a_i} \simeq \sqrt{2} M$, whereas the effective mass of the
inflaton field is $m_\phi \simeq gM/\lambda \ll m_{a_i}$; hence, the
decay of $\phi$-particles into $a_i$-particles is kinematically
forbidden\cite{GarciaBellido:1997wm}.

On the other hand, the decay of $\phi$-particles into $a_r$-particles
is possible because the latter field is effectively massless. Because
the minimum of the potential is at $\phi=0$, an estimate of the decay
width $\Gamma$ is
\begin{equation}
  \Gamma (\phi \phi \to a_r a_r) \sim \frac{(g^4+\delta^2) \phi^2_-}{8\pi
    \bar{m}_\phi} \Phi^2 \sim g^2 \lambda \Phi^2 \,
  . \label{eq:Gamma1}
\end{equation}
However, this exactly resembles the case of incomplete reheating
because $\Phi^2 \sim t^{-2}$, but we would require that $\Gamma$ in
Eq.~(\ref{eq:Gamma1}) to decrease more slowly that $t^{-1}$
\cite{Kofman:1997yn}. This is certainly not the case, and we must
conclude that the inflaton field $\phi$ cannot either transfer its
energy to the field $a_r$.

Successful reheating can be achieved only if the inflaton field
couples to other fields. We can exclude a coupling to other massless
scalar fields because of the same reasons we gave before for the $a_r$
field. Thus, the only way out is to have a coupling of the inflaton
field to (light) fermions $\psi$ with mass $m_\psi < m_\phi$.

For that, we assume that the interaction part of the potential has a
term of the form
\begin{equation}
  \mathcal{L}_{\phi \psi}= h_1 \bar{\psi}\psi \phi \,
  , \label{eq:phipsi}
\end{equation}
where $h_1$ is the interaction constant and is small enough to avoid
large loop corrections to the inflaton potential. The corresponding
rate of decay is\cite{Kofman:1997yn}
\begin{equation}
  \label{eq:inf-decay}
  \Gamma_\phi = \frac{h^2_1 m_\phi}{8 \pi \lambda} \simeq \frac{h^2_1
    g M}{8 \pi \lambda} \, .
\end{equation}

Following standard calculations, the inflaton field decays completely
and the reheating temperature is estimated to
be\cite{Kolb:1990vq,Kofman:1997yn}
\begin{equation}
  \label{eq:reh}
  T_{reh} \simeq 0.1 \sqrt{\Gamma_{\phi} m_{Pl}} \simeq 0.1 h_1
  \sqrt{(g/\lambda) \, M m_{Pl}} \, .
\end{equation}

\subsection{Asymmetric charge after reheating}
After inflation and during the rapid oscillations of the inflaton
field, the Boltzmann equation for the charge of the waterfall field
$n_a$ after inflation is
\begin{equation}
  \dot{ (R^3 n_a)} = - 2 \delta R^3 \, \phi \, a_r \, a_i \, .
\end{equation}
The classical evolution then points out that the source term on the
r.h.s. is negligible because of the smallness of the asymmetry
parameter $\delta$ and of the field $a_r$. Therefore, we conclude that
there is not further production of a charge asymmetry after the end of
inflation. The charge of the $a$-field is then conserved and dilutes
at the usual rate $n_a \sim R^{-3}$.

The final charge stored in the waterfall field can be estimated from
the solutions of the equations of motion after passing by through the
instability points, see Fig.~\ref{fig:1}. The details of the
calculation can be found in\cite{Delepine:2006rn}, and the final
result is
\begin{equation}
  \label{eq:charge}
  |n_a| \simeq \frac{M^2 m^2}{54 \pi^2 H_0} x^2 e^{-x^2} \, ,
\end{equation}
where variable $x$ is defined as
\begin{equation}
  \label{eq:x}
  x = \frac{3 \pi^{3/2}}{2 \lambda^2 g} \frac{\bar{M}^5}{\bar{m}^2}
  \simeq 2.93 \times 10^{-4} \frac{\lambda}{g^2} \, ,
\end{equation}
with $\bar{M} \equiv M/m_{Pl}$ and $\bar{m} \equiv m/m_{Pl}$ are the
Planck normalized values of the waterfall and inflaton mass terms,
respectively.

The asymmetry in Eq.~(\ref{eq:charge}) is not the one we need to
transfer to leptons, but rather we need to calculate the quantity
$|n_a|/s_{reh}$, where $s_{reh} = (2\pi^2/45) q_{reh} T^3_{reh}$ is the
entropy density generated during the reheating process; here, $q_{reh}
\simeq 10^2$ represents the entropic degrees of freedom at the
reheating temperature $T_{reh}$.

After a straightforward calculation taking Eqs.~(\ref{eq:reh})
and~(\ref{eq:charge}) we find
\begin{equation}
  \label{eq:Y-a}
  \frac{|n_a|}{s_{reh}} \simeq 0.01 \frac{\lambda^{1/2}}{g^{5/2}}
  \frac{\bar{M}^{7/2}}{h^3_1} \, .
\end{equation}
Eq.~(\ref{eq:Y-a}) differs from Eq.~(15) in Ref.\cite{Delepine:2006rn}
because in the latter we assumed prompt reheating after
inflation. Also, for the preferred values of the parameters of the
model the exponential term is of the order of $0.1$.

Interestingly enough, the charge asymmetry in the waterfall field only
depends upon the waterfall mass term and any appearance of the
inflaton mass is naturally taken out of the final expression.

\section{Leptogenesis \label{sec:leptogenesis-}}
We now proceed to study the decay of the waterfall field into a
right-handed neutrino, for which we propose an interaction term in the
Lagrangian of the form
\begin{equation}
  \mathcal{L}_{a N_R}= h_{2} \bar{N}^c_{R}N_{R} \, a \, .
\end{equation}
whose corresponding decay width is $\Gamma_a = h^2_2 M/8\pi$.

We shall impose the condition $\Gamma_\phi > \Gamma_a$ to assure that
the reheating process is finished well before the $a$-charge is
transferred to the right handed neutrino. Explicitly, the condition is
$h_2 < (\sqrt{g/\lambda}) h_1$.

When the $a$-particles begin to disintegrate rapidly at a time $t_a =
\Gamma_a^{-1}$, we have a Universe dominated by relativistic fermions
and the dominant particle processes are $CP$ conserving. These are: $a
\longleftrightarrow N_{R}+N_{R}$ and $a^c \longleftrightarrow
N^{c}_{R}+N^{c}_{R}$, with decay width $\Gamma_a$; $N_{R}
\longleftrightarrow  \Phi + \ell_{L}$, and $N^{c}_{R}
\longleftrightarrow  \Phi^c + \ell^{c}_{L}$, with decay width
$\Gamma_D$; and $N_{R} \longleftrightarrow N^{c}_{R}$ with decay width
$\Gamma_{M}$.

The $\Gamma_M$-term appears because of the presence of the non-zero
vev of the waterfall field in the Yukawa coupling with the
right-handed neutrino; it is an interaction term that converts the
right handed neutrino into its own anti-neutrino.

The decay rates corresponding to the right-handed neutrinos and the
leptons are explicitly given by
\begin{equation}
  \label{eq:decays}
  \Gamma_{D} = \frac{\tilde{m}_{1} M^2_{N_R}}{8\pi v^2} \, , \quad \Gamma_{M} =
  M_{N_R}/ 8\pi \, ,
\end{equation}
 where $\tilde{m}_{1}$ is the effective light neutrino mass if the
 origin of neutrino masses comes from the usual see-saw
 mechanism. Taking the experimental limit on neutrino masses, one
 finds that $\tilde{m}_{1} \propto 10^{-10} \, {\rm GeV}$ for $v=174
 \, {\rm GeV}$\cite{Amsler:2008zzb}.

It should be stressed out that in our model there is no $CP$ violation
in the decay of the right handed neutrino; its interaction with the
waterfall field serves only for the transfer of a leptonic number from
the $a$-charge to the leptons of the Standard Model. However, the mass
term of the right-handed neutrino indeed violates leptonic number, and
it acts as a suppression term of the leptonic charge. This shall be
explained in the next section.

\subsection{Boltzmann equations}
Let us write and study the Boltzmann equations for the different
species in our model. Because of the CPT theorem we have $\Gamma(X
\to Y) = \Gamma(X^c \to Y^c)$ for each decay; besides, for each one of
the species we define the quantity $Y=n/s$, which is the number of
particles in a comoving volume. The Boltzmann equations are then the
usual ones of the specialized
literature\cite{Luty:1992un,Buchmuller:2004tu,Buchmuller:2004nz}.

For the waterfall field $a$, we write
\begin{widetext}
\begin{subequations}
\begin{eqnarray}
  \frac{dY_{a}}{dz} &=& - \frac{z}{sH(z=1)} \left[
    \frac{Y_{a}}{Y_{a}^{eq}} \gamma( a \to N_{R} + N_{R})
    - \frac{Y_{N_{R}}Y_{N_{R}}}{Y_{N_{R}}^{eq}Y_{N_{R}}^{eq}} \gamma(
    N_{R} + N_{R} \to a) \right. \nonumber \\
    & & \left.  +\frac{Y_{a}}{Y_{a}^{eq}} \gamma( a \to \phi  + \phi)
    + \frac{Y_{a}^2}{Y_{a}^{eq2}} \gamma( a +a \to \phi +\phi)
  \right] \, , \\
  \frac{dY_{a^c}}{dz} &=& - \frac{z}{sH(z=1)} \left[
    \frac{Y_{a^c}}{Y_{a^c}^{eq}} \gamma( a^c \to N_{R}^c + N_{R}^c ) -
    \frac{Y_{N_R^c}Y_{N_R^c}}{Y_{N_R^c}^{eq} Y_{N_R^c}^{eq}} \gamma (
    N_{R}^c + N_{R}^c \to a^c) \right. \nonumber \\
    &&\left. + \frac{Y_{a^c}}{Y_{a^c}^{eq}} \gamma( a^c \to \phi  + \phi)
    + \frac{Y_{a^c}^2}{Y_{a^c}^{eq2}} \gamma( a^c +a^c \to \phi
    +\phi) \right] \, ,
\end{eqnarray}
\end{subequations}
\end{widetext}
where $z = m_{N_R}/T$, $H(z_a=1)$ is the Hubble parameter at $T =
M_{N_R}$, $Y_{a} = n_{a}/s$, $Y_{a^c}= n_{a^c}/s$, $Y_{N_R} =
n_{N_R}/s$, and $Y_{N_R^c} = n_{N_R^c}/s$; in all terms, $s$ denotes
the entropy density. Notice that we are assuming that the relevant
mass scale for the Boltzmann equations is the mass of the right-handed
neutrino.

For the right-handed neutrino one has
\begin{widetext}
\begin{subequations}
\begin{eqnarray}
  \frac{dY_{N_R}}{dz} &=& - \frac{z}{s H(z=1)} \left[
    \frac{Y_{N_{R}}Y_{N_{R}}}{Y_{N_{R}}^{eq} Y_{N_{R}}^{eq}} \gamma (
    N_{R} + N_{R} \to a) - \frac{Y_{a}}{Y_{a}^{eq}} \gamma ( a \to
    N_{R} + N_{R}) \right] \nonumber \\
   && - \frac{z}{s H(z=1)} \left[ \frac{Y_{N_{R}}}{Y_{N_{R}}^{eq}}
    \gamma(N_{R} \to \Phi + l_{L}) -
    \frac{Y_{\Phi}Y_{l_{L}}}{Y_{\Phi}^{eq} Y_{l_{L}}^{eq}}
    \gamma( \Phi + l_{L} \to N_{R}) \right] \nonumber \\
   && - \frac{z}{s H(z=1)} \left[ \frac{Y_{N_{R}}}{Y_{N_{R}}^{eq}} \gamma(
    N_{R} \to N^c_{R} ) - \frac{Y_{N_{R}}^c}{Y_{N_{R}^c}^{eq}} \gamma^{eq}
    ( N^c_{R} \to N_{R} ) \right] \, , \\
  \frac{dY_{N_{R}^c}}{dz} &=& -\frac{z}{s H(z=1)} \left[
    \frac{Y_{N_{R}^c} Y_{N_{R}^c}}{Y_{N_{R}^c}^{eq} Y_{N_{R}^c}^{eq}}
    \gamma (N^c_{R} + N^c_{R} \to a^c) - \frac{Y_{a^c}}{Y_{a^c}^{eq}}
    \gamma (a^c \to N^c_{R} + N^c_{R} ) \right] \nonumber \\
  &&
  - \frac{z}{s H(z=1)} \left[ \frac{Y_{N_{R}^c}}{Y_{N_{R}^c}^{eq}}
    \gamma (N^c_{R} \to \Phi^c +l^c_{L} ) - \frac{Y_{\Phi^c}
      Y_{l^c_{L}}}{Y_{\Phi^c}^{eq} Y_{l^c_{L}}^{eq}} \gamma( \Phi^c +
    l^c_{L} \to N^c_{R} ) \right] \nonumber \\
  &&
  - \frac{z}{s H(z=1)} \left[ \frac{Y_{N_{R}^c}}{Y_{N_{R}^c}^{eq}}
    \gamma (N^c_{R} \to N_{R}) - \frac{Y_{N_{R}}}{Y_{N_{R}}^{eq}}
    \gamma (N_{R} \to N^c_{R} ) \right] \, ,
\end{eqnarray}
\end{subequations}
whereas for the doublet we have
\begin{subequations}
\begin{eqnarray}
  \frac{dY_{l}}{dz} &=& \frac{z}{s H(z=1)} \left(
    \frac{Y_{N_R}}{Y^{eq}_{N_R}} \gamma( N_{R} \to l + \Phi ) -
    \frac{Y_{l}Y_{\Phi}}{Y^{eq}_{l}Y^{eq}_{\Phi}} \gamma (l + \Phi
    \to N_{R}) \right) \, , \\
  \frac{dY_{l^c}}{dz} &=& \frac{z}{sH(z=1)} \left(
      \frac{Y_{N_R^c}}{Y^{eq}_{N_R^c}} \gamma (N^c_{R} \to l^c +
      \Phi^c ) - \frac{Y_{l^c}Y_{\Phi^c}}{Y^{eq}_{l^c}Y^{eq}_{\Phi^c}}
      \gamma (l^c + \Phi^c \to N^c_{R}) \right) \, .
\end{eqnarray}
\end{subequations}
\end{widetext}
The terms $\gamma(a \to Y)$ are defined as
\begin{equation}
  \gamma (a \to Y) = n_a^{eq} \, \frac{K_1(z)}{K_2(z)} \Gamma (a \to
  Y) \, ,
\end{equation}
where $K_{1}(z)$ and $K_{2}(z)$ are the modified Bessel functions, and
$\Gamma$ is the usual decay width at zero temperature in the rest
frame of the decaying particle. For two-body scattering we have
\begin{equation}
 \gamma (a + b \to Y) = n_a^{eq} n_b^{eq} \langle \sigma (a + b \to Y)
 \vert v \vert \rangle \, .
\end{equation}
and also
\begin{widetext}
\begin{equation}
  \gamma (aX \to Y) = \int d\pi_a d\pi_X d\pi_Y (2\pi)^4
  \delta^{4}(p_a+p_X-p_Y) f_{a}^{eq} f_{x}^{eq} \vert M(aX \to Y)
  \vert^2 \, , \quad f_a^{eq} = e^{-E_a/T} \, .
\end{equation}
\end{widetext}

\subsection{Solving the Boltzmann Equations}
In order to find semi-analytical solutions of the Boltzmann equations,
we take into account only the dominant terms in each equation as
described in Sec.~\ref{sec:leptogenesis-} and assume the standard
condition of kinematic equilibrium for each species. The resulting
equations are
\begin{subequations}
\label{eq:Boltz-simple}
\begin{eqnarray}
  \frac{dY_{a}}{dz} &=& - \frac{\Gamma_{a}}{H(z=1)} \frac{z
    K_1(z)}{K_2(z)} \left(Y_a - Y^{eq}_a \right) \, , \\
  \frac{dY_{N_R}}{dz} &=& - \frac{\Gamma_{a}}{H(z=1)}
  \frac{zK_1(z)}{K_2(z)} \left( Y_a^{eq} - Y_{a} \right) \nonumber \\
  && - \frac{\Gamma_D}{H(z=1)} \frac{zK_1(z)}{K_2(z)} \left( Y_{N_R} -
    Y^{eq}_{N_ R} \right) \nonumber \\
  && - \frac{\Gamma_M}{H(z=1)} \frac{zK_1(z)}{K_2(z)} \left( Y_{N_R} -
    Y_{N_R^c} \right) \, , \\
  \frac{dY_{l}}{dz} &=& \frac{\Gamma_{D}}{H(z=1)} \frac{zK_1(z)}{K_2(z)}
  \left( Y_{N_R} - Y_{N_R}^{eq} \right) \, .
\end{eqnarray}
\end{subequations}
The equations for the antiparticles are exactly the same because in
our model the $CP$ symmetry is conserved in the true vacuum.

In writing Eqs.~(\ref{eq:Boltz-simple})we made two further
assumptions. First, that the decay of the waterfall field occurs out
of equilibrium such that the inverse process $N_R \to a$ is
kinematically forbidden because $M > M_{N_R}$. Second, the same
reasoning applies for the inverse process $l + \Phi \to N_R$ and then
the mass of the leptons should be smaller than that of the
right-handed neutrino.

It should be noticed that the inverse processes involving the inflaton
have also been neglected everywhere in the Boltzmann equations for the
$a$-field~(\ref{eq:Boltz-simple}) and~(\ref{eq:afield}). This
approximation is well justified because we previously assumed that
$\Gamma_a \ll \Gamma_{\phi}$ and then the inflaton field should have
decayed completely at the time the leptogenesis process is taking
place.

However, we should also prevent the creation of inflaton particles
mediated by the decay rates
\begin{subequations}
  \label{eq:a-phis}
  \begin{eqnarray}
    \Gamma (a \to \phi + \phi) \propto \frac{g^4 M}{64 \pi \lambda^2}
    \ll \Gamma_a \, , \\
    \Gamma (a + a \to \phi + \phi) \propto \frac{g^4 M}{32 \pi} \ll
    \Gamma_a \, ,
  \end{eqnarray}
\end{subequations}
In order to prevent any wash-out of the $a$-asymmetry due to their
interactions with the inflaton field it will suffice to impose the
condition $g^2 \ll h_2$.

Let us define $\Delta_{b} = Y_{b} - Y_{b^c}$ for each species. The
Boltzmann equation for the waterfall field simply reads
\begin{equation}
  \label{eq:afield}
  \frac{d\Delta_a}{dz} = - K_a \frac{z^2}{2+z} \Delta_a \, .
\end{equation}
Its solution with an initial condition $\Delta_a(0) = \Delta_a^0$ is
given by
\begin{equation}
  \label{eq:asolution}
  \Delta_a(z) = \Delta_a^0 \frac{e^{K_a z (4-z)/2 }}{(1+z/2)^{4K_a}}
  \, ,
\end{equation}
where $K_a \equiv \Gamma_{a}/H(z=1)$.

Likewise, the Boltzmann equation for the right-handed neutrino is
\begin{equation}
  \label{eq:nrfield}
  \frac{d\Delta_{N_R}}{dz} = \frac{z^2}{2 +z}(K_a \Delta_a -
  K_{DM} \Delta_{N_R}) \, ,
\end{equation}
where $K_{DM} \equiv (\Gamma_{D} + 2\Gamma_{M} )/ H(z=1)$. Up to
quadratures, its solution under the initial condition
$\Delta_{NR}(z=0) = 0$ is
\begin{equation}
  \label{eq:D-R}
  \Delta_{N_R} = \frac{e^{K_{DM} z (4-z)/2 }}{(1+z/2)^{4K_{DM}}} \int^z_0
  \frac{K_a x^2}{2 + x} \Delta_a(x) \, dx \, .
\end{equation}

Finally, the equation for the leptonic doublet is
\begin{equation}
  \label{eq:lfield}
 \frac{d\Delta_l}{dz} = \frac{z^2}{2+z} K_D \Delta_{N_R} \, ,
\end{equation}
where $K_D \equiv \Gamma_D /H(z=1)$.

The resulting leptonic charge can be obtained from the integration of
Eq.~(\ref{eq:lfield}), but for that we need also the analytic solution
of Eq.~(\ref{eq:D-R}). Fortunately, that is not necessary becasue
Eqs.~(\ref{eq:afield}),~(\ref{eq:nrfield}), and~(\ref{eq:lfield}) can
be combined together to get a single equation for the three
abundances. After integration under the initial conditions
$\Delta_l(z=0) = 0$, it can be shown that
  \begin{equation}
    \label{eq:solution1}
    \Delta_l + \frac{K_D}{K_{DM}} \left( \Delta_{NR} + \Delta_a
    \right) = \frac{K_D}{K_{DM}} \Delta^0_a  \, .
  \end{equation}
Eq.~(\ref{eq:solution1}) is the main result in our work and represents
the evolution and transfer of the $a$-charge through the leptogenesis
process.

\subsection{Lepton and baryon asymmetries}
According to Eq.~(\ref{eq:asolution}), not a piece of the waterfall
field charge survives the transfer process and then $\Delta_a \to 0$
in the limit $z \to \infty$. The same will happen to the charge of
the right-handed neutrino in the same limit, see
Eq.~(\ref{eq:D-R}). Therefore, the only surviving charge will be the
leptonic one and the final expression is
\begin{equation}
  \label{eq:leptosol}
  \Delta^\infty_l = \frac{K_D}{K_{DM}} \Delta^0_a =
  \frac{\Delta_a^0}{1 + 2\Gamma_M / \Gamma_D}\, .
\end{equation}

The lepton asymmetry~(\ref{eq:leptosol}) should be further converted
into a baryon charge
asymmetry\cite{Fukugita:1986hr,Luty:1992un,Riotto:1998bt,Buchmuller:2005eh,Frere:2008ct}. Essentially,
one would have $\Delta_l \sim \Delta_B$, where the proportionality between the
lepton and baryon asymmetries depends upon the particle contents of
the model\cite{Harvey:1990qw,Rubakov:1996vz,Kuzmin:1985mm}.

Taking into account the initial asymmetry of the waterfall field, see
Eq.~(\ref{eq:Y-a}), and also the values of the decay widths given
in Eq.~(\ref{eq:decays}), the baryon number would be approximately
given by
\begin{equation}
\label{eq:final}
  \Delta_B \simeq 0.01 \frac{\lambda^{1/2}}{g^{5/2} h^3_1}
  \frac{\bar{M}^{7/2}}{1 +\frac{2v^2}{\tilde{m}_1 M_{N_R}}} \simeq
  \frac{h^{-3}_1 \bar{M}^{7/2}}{1 +2/ \bar{M}_{N_R}} \, ,
\end{equation}
where we have defined the (dimensionless) mass parameter
$\bar{M}_{N_R} \equiv \tilde{m}_1 M_{N_R}/v^2$, and also considered
that $\lambda =1$ and $g \simeq
0.1$\cite{Delepine:2006rn} in the very last equality.

Interestingly enough, the leptogenesis process results in a final
baryon asymmetry that is just the original one stored initially in the
waterfall field except for a term that involves the mass of the
right-handed neutrino.

The baryon charge asymmetry of the Universe is known to be in the following
range\cite{Amsler:2008zzb}
\begin{equation}
  4 \times 10^{-11} \leq \Delta_B\equiv\frac{n_B}{s} \leq 1.4 \times
  10^{-10} \, . \label{eq:baryon}
\end{equation}

In the case $\bar{M}_{N_R} \geq 2$, we see that the value of the
baryon asymmetry is solely provided by the mass value of the
waterfall field, and then $\bar{M} \simeq h^{6/7}_1
\Delta^{2/7}_B$. Recalling the constraint $g^2 < h_2 <
\sqrt{g/\lambda} \, h_1$ and the one arising from cosmic strings
$\bar{M}^2/\lambda^2 < 10^{-6}$\cite{Delepine:2006rn}, then we find
\begin{equation}
  \label{eq:h-1}
  0.03 \lesssim h_1 \lesssim 1 \, .
\end{equation}

All the parameters appear to be tightly constrained. A simple
possibility is just to set $h_1 = 1$, $h_2 = 0.1$,
$\bar{M}_{N_R} =1$, and then $\bar{M} \simeq \Delta^{2/7}_B \simeq 7
\times 10^{-4}$. In other words, the mass scales would be $M \simeq
7.8 \times 10^{15} \textrm{GeV}$ and $M_{N_R} \simeq 3 \times 10^{14}
\textrm{GeV}$. Of course, larger values for the mass of the
right-handed neutrino can also be considered, but we prefer a scenario
in which $M > M_{N_R}$. The reheating temperature, according to
Eq.~(\ref{eq:reh}) is estimated to be $T_{reh} \simeq 9 \times 10^{15}
\textrm{GeV}$.

The other case is to have $\bar{M}_{N_R} \ll 2$, in which the mass
parameter of the right-handed neutrino participates in the final value
of the baryon asymmetry; actually, the baryon asymmetry would now
read $\Delta_B \simeq h^{-3}_1 \bar{M}^{7/2} \bar{M}_{N_R}$.

It is clear that the appearance of the right-handed neutrino mass asks
for larger values of $\bar{M}$ in order to accomplish the baryon
constraint~(\ref{eq:baryon}) . But, as we have seen in the simple
exercise above, the cosmic strings constraint does not support large
values of $\bar{M}$, and then the case $\bar{M}_{N_R} \ll 2$ seems to
be the only one allowed.

\section{Final remarks}
We have studied the transfer of an initial $a$-charge asymmetry
produced at the end of inflation and stored in a complex waterfall
field into a lepton asymmetry. In order to transfer this initial
charge asymmetry first into a lepton asymmetry, and then to a baryon
asymmetry, the model should fulfill two conditions.

First, one needs a mechanism of efficient reheating to produce enough
waterfall particles; second, the waterfall field must be coupled to
leptons. In order to solve the reheating process, we had to introduce
new fermions which couple only to the inflaton field. These fermions
may be candidates for dark matter as we further assumed they should
have very weak interactions with Standard Model particles. As for the
second condition, we assumed that the same waterfall field is the
origin of the Majorana masses for the right-handed neutrinos.

Different constraints entered into play at each one of the stages of
the model, but we were able to show that the simplest realization does
not entail unnatural values for the diverse parameters. For instance,
it was not necessary to introduce new fields in the leptogenesis
process apart from the usual ones in the literature, nor the coupling
parameters were obliged to have embarrassingly small values.

Once the $a$-charge is transformed into a baryon asymmetry we do not
expect further changes because Standard Model interactions preserve
the $B-L$-quantum number. It is important to stress out that our
proposal is different to other leptogenesis models because in our
approach there is not need for $CP$ violating phases in the leptonic
sector.

\begin{acknowledgments}
C.M. thanks the Departamento de F\'isica of the Universidad de
Guanajuato for its kind hospitality in a stay where part of this work
was done. This work has been supported by CONACYT (56946), DINPO and
PROMEP projects.
\end{acknowledgments}

\bibliography{leptorefs}
\end{document}